\documentclass[sigconf,balance=false]{acmart}
\usepackage{popets}
\setcopyright{popets}
\copyrightyear{YYYY}
\acmYear{YYYY}
\acmVolume{YYYY}
\acmNumber{X}
\acmDOI{XXXXXXX.XXXXXXX}
\acmISBN{}
\acmConference{Proceedings on Privacy Enhancing Technologies}
\usepackage[T1]{fontenc}
\usepackage{graphicx}
\usepackage{subcaption}
\usepackage{xcolor}
\definecolor{green}{RGB}{0, 150, 0}
\usepackage{popets}
\usepackage{algpseudocode}
\usepackage{amsmath}
\usepackage[linesnumbered,ruled,vlined]{algorithm2e}
\usepackage{enumitem}
\usepackage{blindtext}
\usepackage{tabulary}
\usepackage{float}
\usepackage{booktabs}
\usepackage{tabularx}
\usepackage{caption}
\begin{document}
\title[ACCESS-FL]{ACCESS-FL: Agile Communication and Computation for Efficient Secure Aggregation in Stable Federated Learning Networks}





\author{Niousha Nazemi}
\affiliation{
  \institution{School of Computer Science,}
\country{University of
  Sydney, Australia}
}
\email{niousha.nazemi@sydney.edu.au}

\author{Omid Tavallaie}
\affiliation{%
  \institution{Department of Engineering Science,}
 \country{University of
  Oxford, UK}
}
\email{omid.tavallaie@eng.ox.ac.uk}

\author{Shuaijun Chen}
\affiliation{%
  \institution{School of Computer Science,}
\country{University of Sydney, Australia}
}
\email{shuaijun.chen@sydney.edu.au}

\author{Anna Maria Mandalari}
\affiliation{%
  \institution{Electronic \& Electrical Engineering,}
\country{University College London, UK}
}
\email{a.mandalari@ucl.ac.uk}

\author{Kanchana Thilakarathna}
\affiliation{%
  \institution{School of Computer Science,}
\country{University of
  Sydney, Australia}
}
\email{Kanchana.thilakarathna@sydney.edu.au}

\author{Ralph Holz}
\affiliation{%
  \institution{Department of Computer Science,}
\country{University of Münster, Germany}
}
\email{ralph.holz@uni-muenster.de}

\author{Hamed Haddadi}
\affiliation{%
  \institution{Department of Computing,}
\country{Imperial College London, UK}
}
\email{h.haddadi@imperial.ac.uk}

\author{Albert Y. Zomaya}
\affiliation{%
  \institution{School of Computer Science,}
 \country{University of
  Sydney, Australia}
}
\email{albert.zomaya@sydney.edu.au}

\renewcommand{\shortauthors}{N. Nazemi, O. Tavallaie, S. J. Chen, A. M. Mandalario, K. Thilakarathna, R. Holz, H. Haddadi, A. Y. Zomaya}

\begin{abstract}
Federated Learning (FL) is a promising distributed learning framework designed for privacy-aware applications. FL trains models on client devices without sharing the client's data and generates a global model on a server by aggregating model updates. Traditional FL approaches risk exposing sensitive client data when plain model updates are transmitted to the server, making them vulnerable to security threats such as model inversion attacks where the server can infer the client's original training data from monitoring the changes of the trained model in different rounds. Google's Secure Aggregation (SecAgg) protocol addresses this threat by employing a double-masking technique, secret sharing, and cryptography computations in honest-but-curious and adversarial scenarios with client dropouts. However, in scenarios without the presence of an active adversary, the computational and communication cost of SecAgg significantly increases by growing the number of clients. To address this issue, in this paper, we propose ACCESS-FL, a communication-and-computation-efficient secure aggregation method designed for honest-but-curious scenarios in stable FL networks with a limited rate of client dropout. ACCESS-FL reduces the computation/communication cost to a constant level (independent of the network size) by generating shared secrets between only two clients and eliminating the need for double masking, secret sharing, and cryptography computations. To evaluate the performance of ACCESS-FL, we conduct experiments using the MNIST, FMNIST, and CIFAR datasets to verify the performance of our proposed method. The evaluation results demonstrate that our proposed method significantly reduces computation and communication overhead compared to state-of-the-art methods, SecAgg and SecAgg+.
\end{abstract}


{\small
\maketitle
}

\section{Introduction} \label{Introduction}
Federated Learning (FL) \cite{mcmahan2017communication} has emerged as a promising approach for privacy-preserving collaborative learning, enabling multiple parties to train machine learning models without sharing their sensitive data. FL allows distributed clients to collaboratively train a model while keeping their data decentralized and private. In FL, each client trains a global model by using its local dataset. Then. it sends the trained model update to the central server, which builds a new global model by aggregating all updates. While FL protects user privacy by avoiding direct data sharing, it still faces challenges and vulnerabilities \cite{kairouz2021advances}, such as model inversion attacks \cite{fredrikson2015model}, where an honest-but-curious server \cite{kissner2005privacy, olson2007harvesting} may reconstruct the original client data by reverse-engineering the local model weights.
\begin{figure}[t]
  \centering
    \includegraphics[width=80 mm, height=33 mm]{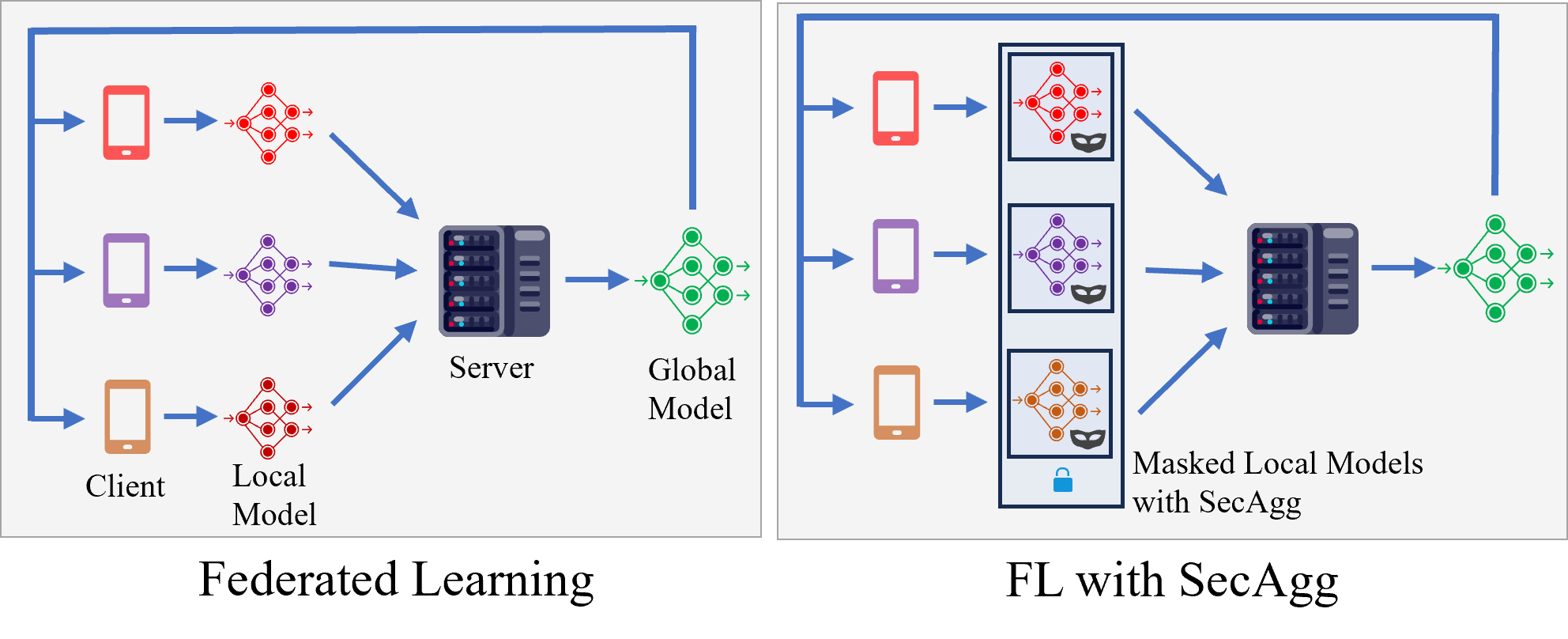} \vspace{-4 mm} %
  \caption{Comparison between vanilla FL and FL with SecAgg.} \vspace{-7 mm}
  \label{fig:comparision_fl_SecAgg}
\end{figure}
To address these privacy threats, Google proposed the Secure Aggregation (SecAgg) protocol \cite{bonawitz2017practical} as a secure multi-party computation (MPC) \cite{cramer2015secure} method based on Diffie-Hellman (DH) key agreement \cite{diffie1976new} and Shamir's secret sharing \cite{dawson1994breadth, pang2005new}. In SecAgg, each client generates a shared secret for every other client participating in a training round. This secret is used in a Pseudo-Random Generator (PRG) function \cite{blum1986simple, impagliazzo1989pseudo} to create a mask that is added to the model update, concealing individual updates during transmission and preserving privacy through a double-masking technique. This technique, which involves shared masks between clients and generating a self-mask for each client, ensures that the server only learns the aggregated result rather than the model weights of individual clients. SecAgg also handles client dropouts and message delays by reconstructing the shared mask of a dropped client, and self-mask of participating clients. Figure \ref{fig:comparision_fl_SecAgg} illustrates the SecAgg approach in comparison to the traditional FL scheme. 

SecAgg is effective against active adversaries \cite{kissner2005privacy, miller2014adversarial}, where malicious clients and the server actively collude to determine masks added to model updates to infer sensitive information about the participating clients' private data. However, as the network size grows, SecAgg incurs high computation costs. Each client must perform cryptographic operations and key generation for every other client in the network. Moreover, the server's communication overhead increases as it needs to reconstruct masks for dropped clients. To address these limitations, Google proposed SecAgg+ \cite{SecAgg+}, an improved version of SecAgg. In SecAgg+, instead of generating shared secrets with every other client, the server generates a random $k$-regular graph for $k = log(|C|)$ with $|C|$ clients. The clients then generate shared masks with their neighbors in the graph. Although SecAgg+ reduces the communication and computation costs compared to SecAgg, it still causes unnecessary costs in stable networks and is more suitable for unstable networks with frequent client dropouts. In both SecAgg and SecAgg+, the server needs to reconstruct the self-masks of participants. Even when there are no client dropouts and shared masks cancel out automatically, the server still needs to remove the self-masks of participants, which adds unnecessary overhead in stable networks with infrequent client dropouts.

\begin{figure}[t]
  \centering
    \includegraphics[width=80 mm, height=35 mm]{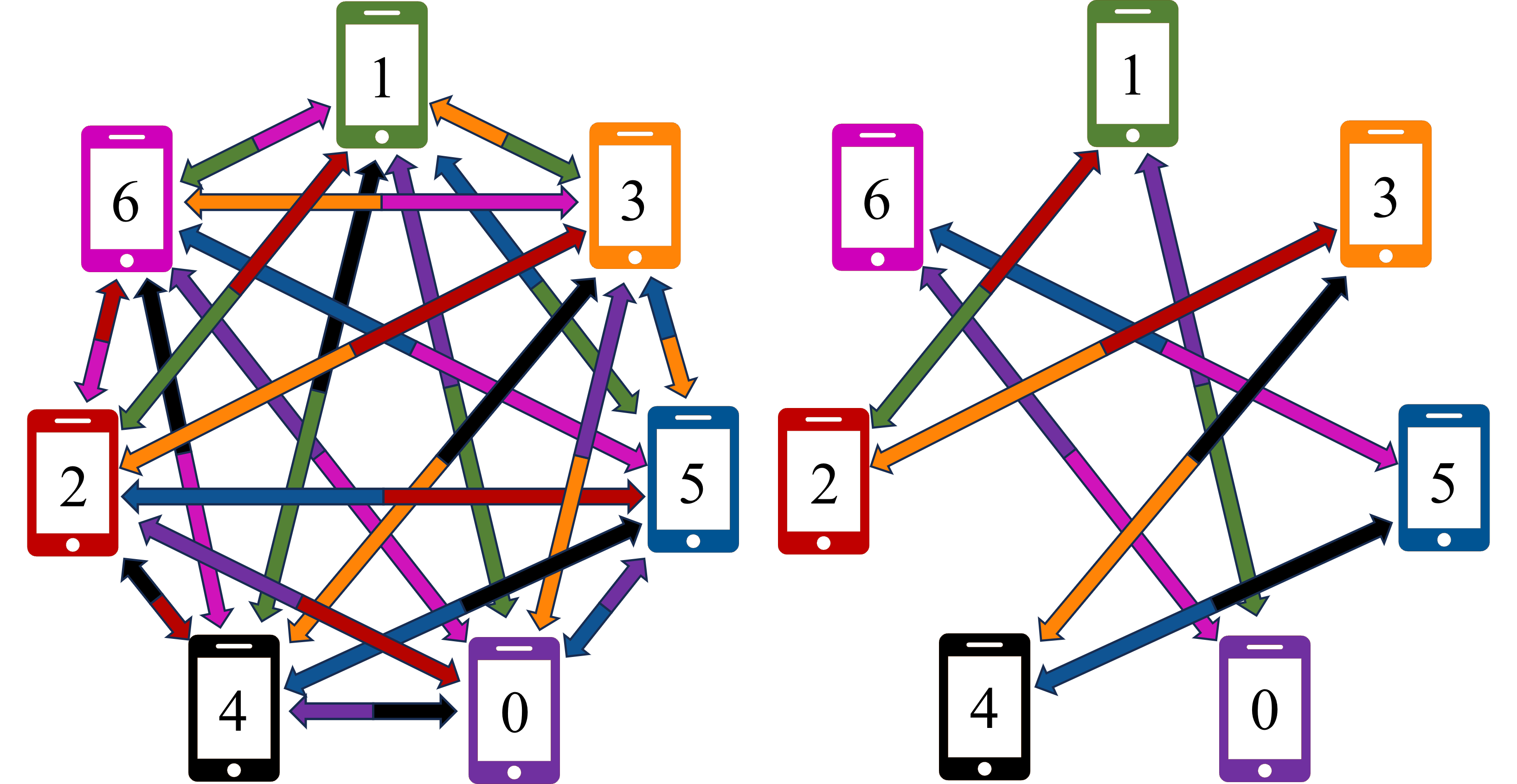}
  \caption{ SecAgg (left) versus ACCESS-FL (right) in finding pairs and creating shared secrets.\vspace{-3 mm}}   
  \label{fig:SecAgg_vs_ACCESS-FL}
\end{figure}

To design an efficient secure aggregation mechanism suitable for FL systems with stable network conditions, in this paper, we propose ACCESS-FL, an enhanced secure aggregation protocol designed based on the key agreement. ACCESS-FL is designed for honest-but-curious FL scenarios with stable network conditions, such as fraud detection for financial applications \cite{yang2019ffd}, privacy-preserving systems against money laundry by IBM \cite{ibm2023privacy}, and AI applications in healthcare systems \cite{rahman2023federated}. In these applications, the network exhibits low delay variations and limited client dropouts. The main contributions of this paper are: \vspace{-1mm}
\begin{itemize} [left=0pt]
    \item By generating shared secrets between only two client devices regardless of the network size, we significantly reduce the communication and computation overhead of ACCESS-FL compared to state-of-the-art methods in honest-but-curious FL scenarios with stable network conditions. ACCESS-FL eliminates the need for each client to perform cryptographic operations and key generation for every other client in the network.
    \item We introduce a dynamic client pairing mechanism based on a deterministic function and a secret seed, ensuring that the pairing is unknown to the server to enhance data privacy. (Figure \ref{fig:SecAgg_vs_ACCESS-FL}).
    \item We simplify the secure aggregation process in ACCESS-FL by eliminating the double-masking technique and the associated cryptographic computations while maintaining the same level of communication as in traditional federated learning for the server, with additional communication only in the case of client dropouts.
\end{itemize} \vspace{-1mm}

Our experiments on the MNIST dataset \cite{lecun1998gradient}, Fashion-MNIST \cite{tensorflow2024fashionmnist} and CIFAR10 \cite{tensorflow2024cifar10} demonstrate that ACCESS-FL significantly reduces communication and computational costs for both clients and the server while maintaining the same level of security against model inversion attacks as SecAgg and SecAgg+ in honest-but-curious FL scenarios with stable network conditions and the limited number of client dropout. The implementation and the source code for ACCESS-FL are publicly available on \cite{accessFL_github}.
The rest of this paper is organized as follows. Section \ref{sec: Preliminary Study: SecAgg and SecAgg+} reviews related work on secure aggregation in FL. Section \ref{sec: Security Primitives} explains the security concepts used in ACCESS-FL. Section \ref{sec: Protocol Overview} describes the ACCESS-FL protocol in detail. Section \ref{sec: Proof of Maintaining Aggregation Result Equal to Traditional FL} presents the theoretical analysis of the correctness and security of ACCESS-FL. Section \ref{sec: Evaluation Result} reports the experimental evaluation of ACCESS-FL in comparison to SecAgg and SecAgg+. Section \ref{sec: Discussion and Future Work} discusses the limitations of ACCESS-FL and future suggestions. Finally, Section \ref{sec: Conclusion} concludes the paper and summarizes our contributions.

\section{Preliminary Study: SecAgg and SecAgg+} \label{sec: Preliminary Study: SecAgg and SecAgg+}

In this section, we provide an overview of Google's Secure Aggregation (SecAgg) protocol and its improved variant, SecAgg+. We explain the process of message passing and discuss its challenges in the context of hones-but-curious FL scenarios with stable network conditions, where client dropouts are limited and delay variations are low. Table \ref{tab:main_notation} summarizes the main notations used throughout this paper. \vspace{-2 mm}

\subsection{Message Passing in SecAgg}
The section explains the SecAgg protocol as follows:
\begin{itemize}[label={}, leftmargin=0pt]

    \item (1) \textbf{Broadcasting the global model:} The server broadcasts the initial global model to clients. 

    \item (2) \textbf{Key pair generation:} Each client ${i}$ generates \textbf{two} private-public key pairs as $(SK_{i}^1, PK_{i}^1)$ and $(SK_{i}^2, PK_{i}^2)$. Then, it sends its public keys to the server.

    \item (3) \textbf{Broadcasting public keys:} The server broadcasts public keys to all clients.

    \item (4) \textbf{Client-side preparation:} Each client $i \in C$ generates a random element $b_{i}$, then divides $b_{i}$ and $SK_{i}^1$ into $\left|C\right|$ parts and assigns each part to a client pair $j \in C$ ($b_{i,j}$, $SK_{i,j}^1$). Then, Client $i$ encrypts a message $(i || j || b_{i,j} || SK_{i,j}^1)$ for each pair $j$ (by using a key generated from $SK_{i}^2$ and $PK_{j}^2$) to create a cipher text $e_{i,j}$. Finally, the client sends all generated $e_{i,j}$ to the server.

    \item (5) \textbf{Distribution of cipher texts:} The server collects these cipher texts and puts participating clients in the $C_{1}$ set. Here, we assume that $\left|C_{1}\right| = \left|C\right|$, hence the server sends $\left(|C| - 1\right)$ encrypted values to every client. 

    \item (6) \textbf{Masked model generation:} Each client $i$ creates $\left(|C| - 1\right)$ shared secrets with every other client $j$ by using $SK_{i}^1$ and $PK_{j}^1$. Then, client $i$ expands these created shared secrets and its random element $b_{i}$ by the pseudo-random generator function $PRG$ to create a self-mask $m_{i}$ and shared masks $m_{i,j} \; \forall j \in C_{1}$. By using these masks, client $i$ computes the masked model $w_{i}^{mask}$ from its trained model $w_{i}$, which is sent to the server. 

    \item (7) \textbf{Participants awareness:} The server creates a set $C_{2}$ from clients that sent their masked models. Then, the server sends the set $C_{2}$ to all the clients $j \in C_{2}$. Then, each client $i$ identifies the participants and decrypts the received encrypted values $e_{j, i}$ by using a key generated from $SK_{i}^2$ and $PK_{j}^2$. Thus, client $i$ obtains $b_{j,i} \; \forall j \in C_{2}$ for participants and $m_{j, i} \; \forall j \in C_{1} \setminus C_{2}$ for dropped-out clients, and sends them to the server. 

    \item (8) \textbf{Global model aggregation:} The server gathers $\left(|C| - 1\right)$ portions of random elements of participants and dropped-out clients, then expands each reconstructed value by $PRG$ to generate self-mask $m_{j} \; \forall j \in C_{2}$ and shared masks $m_{j, i} \; \forall j \in C_{1} \setminus C_{2}$. Finally, it aggregates the global model by 
    \[
    \sum_{i \in \{C_{2}\}} w_{i}^{masked} - \sum_{i \in \{C_{2}\}} m_{i} + \sum_{i \in C_{2}, j \in \{C_{1} \setminus C_{2}\}} m_{j,i}.
    \]

\end{itemize}

Based on \cite{bonawitz2017practical}, in SecAgg, the communication cost for a client and the server are $O(|C|)$ and $O(|C|^2)$, respectively, where $|C|$ is the number of participating clients.\vspace{-2 mm}

\subsection{Challenges of SecAgg in Stable FL}
\begin{table}[!tbp]
\centering
\caption{Declaration of main notations} \vspace{-2mm}
\begin{tabular}{|p{0.15\columnwidth}|p{0.75\columnwidth}|} \hline
\multicolumn{1}{|c|}{\textbf{Notations}} & \multicolumn{1}{c|}{\textbf{Definition}} \\\hline
$n$                  & Round number of FL \\ \hline
$C$                  & The list of participating clients in an FL round \\ \hline
$|C|$                & Number of clients in the FL system \\  \hline
$param$              & Public parameters for key pair generation \\ \hline
$SK_{i}$             & Private key of client $i$ \\ \hline
$PK_{i}$             & Public key of client $i$ \\ \hline
$PK_{all}$           & List of public keys \\ \hline
$PRG$                & Pseudo Random Generator function \\ \hline     
$s_{i,j}$            & Shared secret generated by $SK_{i}$ and $PK_{j}$ \\ \hline
$fp_{i}$             & First pair of client $i$ for generating shared secret\\ \hline
$sp_{i}$             & Second pair of client $i$ for generating shared secret \\ \hline
$m_{i,j}$            & Shared mask between client $i$ and client $j$ \\ \hline
$w_{i}$              & Trained model of client $i$  \\ \hline
$w_{i}^{masked}$     & Masked model of client $i$ \\ \hline
$W_{aggregated}^{masked}$         & Aggregation of all masked models \\ \hline
$G$                   & Global model \\ \hline
\end{tabular}
\label{tab:main_notation}
\end{table}

\noindent \textbf{Handing Client Dropout or Delayed Messages:}
In a practical FL scenario, issues such as the unstable Internet connection can interrupt the process of creating the global model. In SecAgg, Google uses a double-masking technique to ensure that each client's model updates remain secure against model inversion attacks, even in cases of user dropout or delayed updates. However, this involves several cryptographic operations, including two key pairs generation (public and private keys for creating shared secrets and secure communication), creating shared and self secrets, utilizing Shamir's secret sharing (a method for dividing a secret into parts), conducting encryption and decryption operations, and calling a pseudo-random generator function for generating the masks. All mentioned steps can significantly increase the computational and communication costs, even in stable networks with low dropout rates. For example, in a healthcare FL system, where patient data is being used for training, an unstable connection can cause delays. While the double-masking technique ensures privacy, it significantly increases computational and communication overhead, and each client must compute extensive cryptographic operations. 
 
\noindent \textbf{High computation cost:} 
In SecAgg, each of the $|C|$ clients generates $|C|-1$ shared secrets regarding every other client devices and creates a unique random element for itself. These values are then expanded using a pseudo-random generator function to create shared and individual masks. This process significantly increases the computational complexity of the client device to $O(|C|^2)$ which becomes particularly challenging in large FL systems such as Google Gboard with one billion clients. The server also has substantial computation overhead in SegAgg, as it must reconstruct shared masks for dropped-out clients and regenerate self-masks for participating clients (those who have sent their masked updates). These tasks elevate the server's computational cost to $O(|C|^2)$.

\noindent \textbf{High communication cost:} 
The communication cost for each client in SecAgg is $O(|C|)$ due to sending two public keys and encrypted shares of both the private key and the random element, along with their masked model updates. The server has a higher communication cost, as it is responsible for distributing encrypted values to clients and broadcasting public aggregated model updates. This leads to a quadratic increase in communication cost ($O(|C|^2)$) for the server. In large-scale FL scenarios, such as smart city applications with thousands of participating devices, the server's communication load becomes a significant bottleneck. \vspace{-1mm}

\subsection{SecAgg+}
SecAgg+ \cite{SecAgg+} is an improvement over SecAgg designed to reduce the computational and communication costs associated with secure aggregation. Instead of generating shared secrets regarding every client, the server generates a random $k$-regular graph for $k = (\log_{}{|C|})$, where $|C|$ is the number of clients. Clients only generate shared masks with their neighbors in this graph. While SecAgg+ reduces the costs compared to SecAgg, for a much larger number of clients (e.g., billions), it still leads to unnecessary overhead in stable networks with low rates of client dropouts. \vspace{-3 mm}

\subsection{Advantages of ACCESS-FL}
While SecAgg and SecAgg+ provide secure aggregation mechanisms for FL, they face challenges in terms of high computation and communication costs in honest-but-curious FL scenarios with stable network environments. Our proposed ACCESS-FL protocol aims to address these limitations by \textbf{1)} reducing the number of shared masks to only \textbf{two masks per client regardless of the network's size}, \textbf{2) eliminating the need for executing cryptographic operations with high computational cost} such as encryption, decryption, or Shamir's secret sharing on client devices, \textbf{3) eliminating the need for sharing any values other than one public key and the masked model on client devices} (which leads to a substantial decrease in the number of messages transmitted through the network), and \textbf{4) reducing the server’s computational cost by removing the need for handling mask cancellation}.\vspace{-3 mm}
 
\section{Security Primitives} \label{sec: Security Primitives}
This section introduces the fundamental cryptography used in ACCESS-FL, including the pseudo-random generator function and the key agreement protocol:\\
\noindent \textbf{The Pseudo-Random Generator function (PRG)} \cite{blum1986simple, impagliazzo1989pseudo} is a deterministic function that produces a sequence of outputs that appear random from a given seed input. PRGs are crucial in secure aggregation, as they allow for the generation of masks that hide the individual model updates. In ACCESS-FL, we implement the PRG using Advanced Encryption Standards (AES) \cite{daemen1999aes, rijmen2001advanced} in counter mode (CTR \cite{lipmaa2000ctr}). AES-CTR combines a counter value with a nonce to produce unique inputs for the AES encryption function, resulting in a stream of pseudo-random bits. The counter is incremented for each encryption operation and ensures that the same seed always produces the same pseudo-random sequence. This property allows the PRG to securely expand shared secrets into masks that are compatible with the dimensions of model updates.

\noindent \textbf{Key Agreement Protocols} \cite{ blake1997key}, such as Diffie-Hellman (DH) \cite{diffie2022new} and Elliptic Curve Diffie-Hellman (ECDH) \cite{haakegaard2015elliptic}, enables two parties to securely establish a shared secret \cite{bao2003variations} through following steps:\textbf{(I) Generating public parameters:} A trusted third party generates the public parameters using a function $\textbf{param\_gen} (key\_size) \rightarrow param$. These parameters include a large prime number $p$ and a generator $g$ modulo $p$. Public parameters are shared between both parties and serve as the foundation for the key agreement protocol. \textbf{(II) Generating key pairs:} Using the public parameters, each party generates a key pair consisting of a private key $SK$ (randomly chosen from $[1, p-1]$) and a corresponding public key $PK \equiv g^{SK} \pmod{p}$ through a function $\textbf{key\_gen} (param) \rightarrow key\_pair$. Despite using the same public parameters, the key pairs generated by both parties are unique to each individual. The private key is kept confidential by its owner, while the public key is shared with the other party. In secure aggregation, the server broadcasts each client's public key to every other client. \textbf{(III) Creating Shared Secret:} Each party computes the shared secret using its own private key and the other party's public key. Due to the mathematical properties of the key agreement protocol, both parties arrive at the same shared secret value, which can be used as input for the PRG to generate shared masks. Client $i$ computes the shared secret $s_{i,j}$ as\vspace{-2 mm}
\begin{equation}
s_{i,j}= PK_j^{SK_i} = (g^{SK_j})^{SK_i} = g^{SK_i SK_j} \pmod{p},
\end{equation}
which is equal to the shared secret $s_{j,i}$ computed by the client $j$ as
\begin{equation}
s_{j,i} \equiv PK_i^{SK_j} \equiv (g^{SK_i})^{SK_j} \equiv g^{SK_i SK_j} \pmod{p}.
\end{equation}
Hence, clients $i$ and $j$ compute the same shared secret, which is $\textbf{key\_agree} (SK\_current\_client, PK\_peer) \rightarrow shared\_secret$. In both SecAgg and ACCESS-FL, a key agreement protocol is used to generate shared secrets that serve as inputs for the PRG function. In SecAgg, Google also applies DH to encrypt fragments resulting from Shamir's secret sharing, a step that is not present in our protocol. 
By employing PRGs and key agreement protocols, ACCESS-FL ensures that the individual model updates of clients remain hidden during the aggregation process to protect the privacy of the clients participate in training.

\begin{algorithm}[t]
\caption{Client-side algorithm in ACCESS-FL to generate a key pair (in the first training round).}
\label{alg:client_initial}
\SetAlgoLined
\SetKwFunction{KeyGen}{key\_gen}
Wait for the server to send the initial $G$; \\
Wait for a trusted third party to send public parameters; \\
{\textcolor{green}{\# Client $i$ generates a key pair with public parameters}} \\
$(SK_{i}, PK_{i}) \gets$ \KeyGen{param}; \\
Store $SK_i$ securely;\\
Send $PK_i$ to the server;\\
Wait to receive $PK_{all}$  from the server;\\
\end{algorithm}\vspace{-2 mm}

\section{ACCESS-FL Protocol} \label{sec: Protocol Overview}
In this section, we introduce ACCESS-FL, our proposed communication and computation-efficient secure aggregation protocol for generating masked models during the FL process. ACCESS-FL is designed for stable networks with limited client dropouts and low delay variations in honest-but-curious scenarios. The main stages of ACCESS-FL are: 1) initialization, 2) pairs selection, 3) generating shared masks, 4) local training and mask application, and 5) updating the global model. In the following, we explain details of each stage.

\noindent \textbf{1) Initialization:}
First, the server broadcasts the initial global model to all clients. Each client receives common public parameters from a \textit{trusted third party}; these public parameters are generated through the key agreement algorithm with a specified key size. Then, each client generates a unique public-private key pair using these public parameters through the DH key generation function and sends the public key to the server. The server then broadcasts the list of public keys to all clients. \textbf{ In ACCESS-FL, each client generates only one key pair once through all FL rounds (algorithm \ref{alg:client_initial})}. However, in SecAgg, each client needs to create two key pairs in every FL training round. This is because, in SecAgg, the server reconstructs the self-mask of participants and the shared masks of dropped-out and delayed clients. If a client is delayed, the server in the current round computes its shared masks, and in the next round, if the client remains a participant, the server reconstructs its self-mask. If the key pairs used to generate the shared masks of this client do not change in the next round, the server can possess both the shared masks and self-mask of this client, allowing it to calculate the client's trained model. However, in ACCESS-FL, the server is not responsible for removing masks from the aggregated function. Thus, the server cannot recompute the shared masks of the clients, and there is no need to generate new key pairs in every round. Consequently, \textbf{the server broadcasts the public keys once in ACCESS-FL for all FL rounds}. In contrast, in SecAgg, the server needs to broadcast the newly generated public keys in every round. Therefore, computation and communication costs associated with key pair generation and distribution are significantly reduced in ACCESS-FL. 

\noindent \textbf{2) Pairs Selection:} Upon receiving the public keys, each client identifies two peers to create shared secrets with them. These shared secrets are constructed using the peer's public key and the client's own private key. In ACCESS-FL, clients are sorted into a participating list. Hence, each client has a position as an index in this list. \textbf{To determine the pair's position, each client uses a deterministic function provided by a trusted third party. This function generates a random integer within the range $[1, \lfloor \frac{|C|-1}{2} \rfloor)]$ based on the training round number} (a variable which is known by all clients), ensuring that the distance varies each round. All clients run this distance generator function and calculate the same distance value, ensuring consistency across the network. The distance value ensures that clients have unique pairs each round. To prevent identical pairs in different rounds, the distance generated in each round is a random value within the given domain, excluding the distances used in previous rounds. Once calculated, this position is added to and subtracted from the client’s own index in the sorted participant list to identify the two corresponding pair clients for the creation of shared secrets. As an example, \textbf{for client $i$, two pair indexes are calculated as ($(i+distance) \mod |C|$) and ($(i-distance+|C|) \mod |C|$), where $|C|$ is the number of clients in the participating list}. This dynamic pair selection process ensures that clients have different pairing partners in each round, enhancing the privacy and security of the protocol. Algorithm  \ref{alg:client_find-pair} shows the process of finding pairs in ACCESS-FL for client $i$.

\begin{algorithm}[t]
\caption{Client-side algorithm in ACCESS-FL to find two pairs (during all training rounds).}
\label{alg:client_find-pair}
\SetAlgoLined
\SetKwFunction{SharedSecretGen}{key\_agree}
\SetKwFunction{PRGFun}{PRG}

{\textcolor{green}{\# Client $i$ calculates the distance value to find its pairs}} \\
$distance_i^n = \text{RandInt}(\text{set}_i^n)$; \\

$set_i^n = \{d \mid \forall d \in [1, \lfloor \frac{|C|-1}{2} \rfloor], d \neq \text{distance}_i^{n-1} \};$ \\

{\textcolor{green}{\# Client $i$ finds its pairs from the sorted participant list}}\\
$fp_{i} \gets (i+distance_i^n) \mod  |C|$; {\textcolor{green}{\# First pair's index}} \\
$sp_{i} \gets (i-distance_i^n+|C|) \mod  |C|$; {\textcolor{green}{\# Second pair's index}} \\
\end{algorithm}

\noindent \textbf{3) Shared Masks Generation:}
After finding pairs, each client uses the private key and the public keys of its peers to generate a shared secret based on the key agreement algorithm. The shared secret, unique to each pair, is then used as the seed for the PRG function to generate shared masks. Based on each client's index in the participant list, the one with a smaller index gets the -1 coefficient added to its shared masks. Since we use the key agreement algorithm, the shared secret generated by the private key of client $i$ and the public key of client $j$ equals the shared secret generated by client $j$'s private key and client $i$'s public key. These shared secrets are then input into the PRG function, which produces identical outputs given the same input. Thus, \textbf{$shared\_mask_{i,j} = shared\_mask_{j,i}$. By using a coefficient of -1 for the client with the smaller index, we ensure that $shared\_mask_{i,j}$ and $shared\_mask_{j,i}$ cancel out each other in the aggregation process}. This approach simplifies the masking process and reduces the computational burden on clients compared to SecAgg's double-masking technique.

\noindent \textbf{4) Local Training and Mask Application:}
Each client trains the global model by using its local dataset. \textbf{ACCESS-FL is designed to be independent of the model and data distribution types, making it versatile for various applications}. After training the model locally, each client applies the masking vectors to its model updates to generate its masked model. The masked model is then sent to the server, ensuring that the server does not have access to the plain-trained model of each client (algorithm \ref{alg:client_masked-model}). This step ensures the privacy of individual clients' models while allowing the server to aggregate the masked models into a global model.

\begin{algorithm}[t]
\caption{Client-side algorithm in ACCESS-FL to generate its masked model (during all training rounds).}
\label{alg:client_masked-model}
\SetAlgoLined

\SetKwFunction{SharedSecretGen}{key\_agree}
\SetKwFunction{PRGFun}{PRG}

{\textcolor{green}{\# Client $i$ generates shared secret with its two pairs.}} \\
$s_{i,fp_{i}} \gets$ \SharedSecretGen{$SK_{i}$, $PK_{fp_{i}}$}; \\
$s_{i,sp_{i}} \gets$ \SharedSecretGen{$SK_{i}$, $PK_{sp_{i}}$}; \\

{\textcolor{green}{\# Client $i$ creates its shared masks through \PRGFun function.}} \\
$m_{i,fp_{i}} \gets$ \PRGFun{$s_{i,fp_{i}}$}; \\
$m_{i,sp_{i}} \gets$ \PRGFun{$s_{i,sp_{i}}$}; \\

{\textcolor{green}{\# Determine signs based on indices}}\\
$sign_{fp} \gets \text{if } fp_{i} < i \text{ then } -1 \text{ else } 1$; \\
$sign_{sp} \gets \text{if } sp_{i} < i \text{ then } -1 \text{ else } 1$; \\

{\textcolor{green}{\# Client $i$ calculates its masked model.}}\\
$w_{i}^{masked} \gets w_{i} + sign_{fp} \times m_{i,fp_{i}} + sign_{sp} \times m_{i,sp_{i}}$; \\
Send $w_{i}^{masked}$ to the server; \\
Wait to receive the new $G$ from the server;
\end{algorithm}

\begin{figure*}[t]
  \centering
    \includegraphics[width=0.85\linewidth]{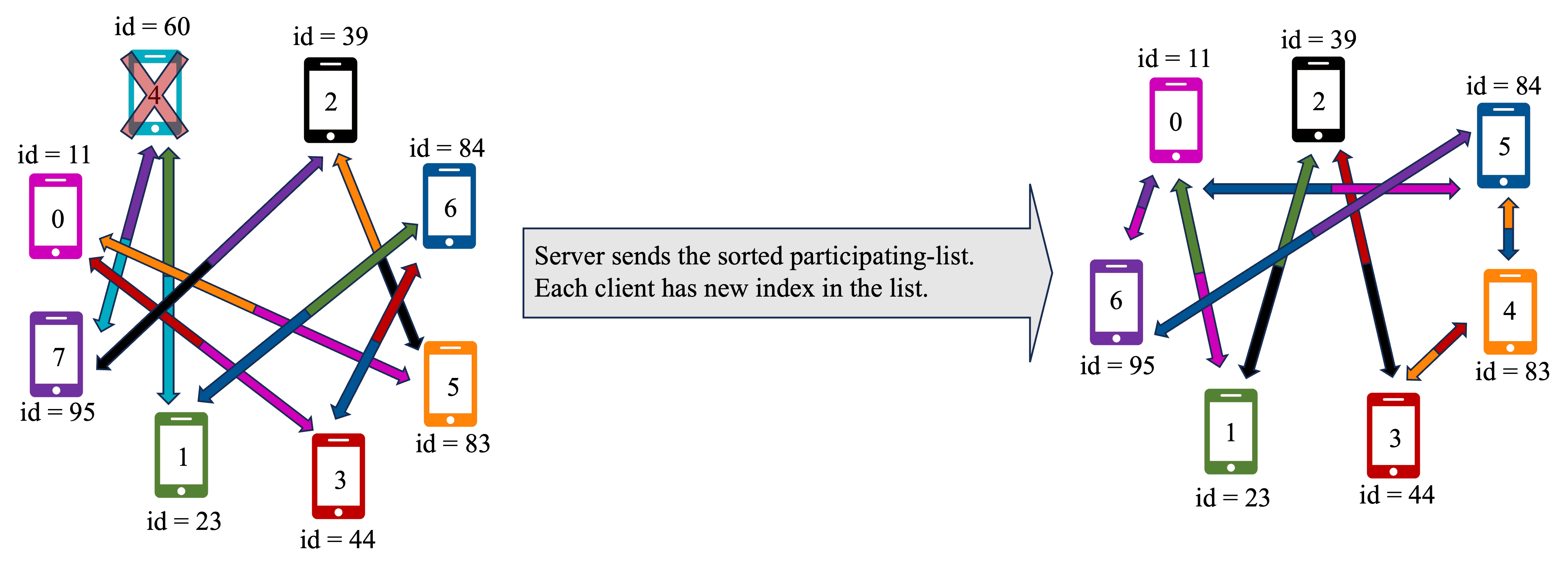}\vspace{-2mm}
  \caption{Finding new pairs in the presence of a client drop-out.}\vspace{-2mm}
  \label{fig:new_pair}
\end{figure*}

\noindent \textbf{5) Global Model Update:}
In this phase, the server waits to receive the masked models from all clients within a specific time frame. If the server gets the masked models from all clients within this period, it aggregates them to generate the new global model. Algorithm \ref{alg:server} shows the process of ACCESS-FL running at the server. However, if a client drops out or the server receives a masked model after the specified time, the server broadcasts the sorted list of participating clients that have sent their masked models and waits for the new masked model from these participating clients. In this scenario, \textbf{all clients recalculate the distance to find new pairs and send their new masked models (algorithm \ref{alg:client_dropout}). This step is mandatory to remove the shared masks associated with the dropout client}, preventing any deviation in the aggregation result.
\begin{algorithm}[t]
\caption{Server-side algorithm in ACCESS-FL.}
\label{alg:server}
\SetAlgoLined
{\textcolor{green}{\# First training round}} \\
Broadcast initial $G$; \\
Wait for all clients to send their public keys (Algorithm \ref{alg:client_initial});\\
\For{$\forall i \in C$}{
    $PK_{all} \cup [PK_{i}]$; {\textcolor{green}{\# List of public keys}}\\
}
Broadcast list of $PK_{all}$; \\

{\textcolor{green} {\# Second training round onwards}} \\
Wait for clients to send their masked models (Algorithm \ref{alg:client_masked-model});\\
\If{number of masked models $< |C|$}{
    {\textcolor{green}{\# Server updates $C$ with participants}}\\
    $C \gets$ list of participants who sent masked models;\\
    {\textcolor{green}{\# Server sends updated $C$ to all clients}}\\
    \For{$\forall i \in C$}{
        Send updated $C$ to client $i$\;
    }
    Wait for clients to send their masked models (Algorithm \ref{alg:client_masked-model});\\
}    
{\textcolor{green}{\# Server aggregates masked models}} \\
$W_{aggregated}^{masked} \gets 0$; \\
$W_{aggregated}^{masked} \gets \sum_{i \in C} w_{i}^{masked}$; {\textcolor{green}{\# Sum of masked models}}\\
$G \gets W_{aggregated}^{masked}$; \\
Broadcast new $G$;
\end{algorithm}
\begin{algorithm}[b]
\caption{Client-side algorithm in ACCESS-FL for handling client dropout or delayed updates.}
\label{alg:client_dropout}
\SetAlgoLined
{\textcolor{green}{\# Client $i$ calculates the $new\_distance_{i}^n$ }} \\
$new\_distance_i^n = \text{RandInt}(\text{set}_i^n)$; \\
$set_i^n = \{d \mid \forall d \in [1, \lfloor (|C| - 1) / 2 \rfloor], d \neq distance_i^{n-1} \And distance_i^{n} \};$ \\
$distance_i^n \gets new\_distance_i^n$; \\
Find new pairs with new distance by Algorithm \ref{alg:client_find-pair};\\
Calculate $w_{i}^{masked}$ by Algorithm \ref{alg:client_masked-model} ;\\
Send $w_{i}^{masked}$ to the server;\\
Wait for the server to send new $G$;\\
\end{algorithm}
The goal is to ensure that the aggregation output is equivalent to traditional FL aggregation output. ACCESS-FL is designed for stable FL environments with limited client dropouts and low delay variations, providing the server does not get stuck in the same training round waiting for new masked models. Figure \ref{fig:new_pair} shows an example where a client dropout occurs among 8 clients, and the participants need to find new pairs to generate new shared masks. By handling client dropouts and delayed updates, ACCESS-FL maintains the integrity of the aggregation process while minimizing the computational overhead. 
\begin{figure*}[h!]
  \centering
    \includegraphics[width=0.85\linewidth]{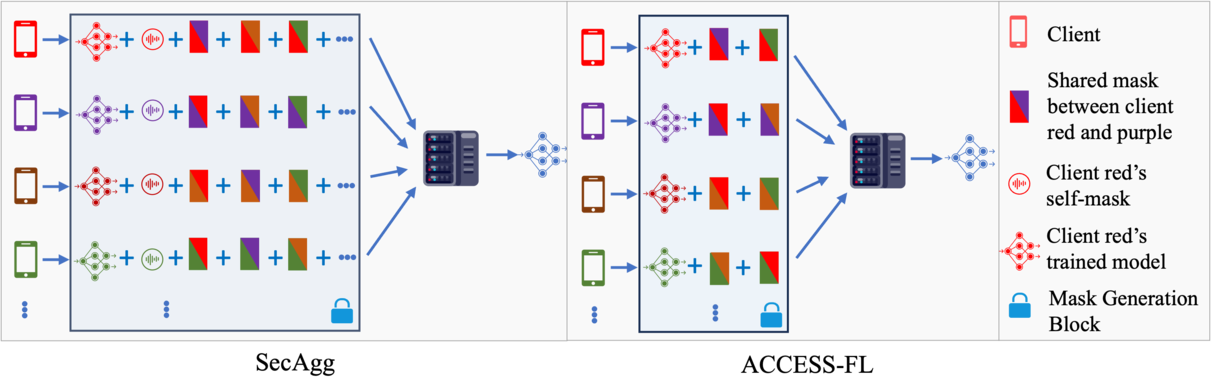}\vspace{-2mm}
  \caption{Comparison between SecAgg and ACCESS-FL.}
  \label{fig:comparison_SecAgg_accessfl} \vspace{-2mm}
\end{figure*}
Through the stages mentioned above, ACCESS-FL facilitates the privacy-preserving aggregation that allows multiple clients to contribute to an aggregated result without exposing their individual trained models while maintaining low communication and computation costs. The diagram in Figure \ref{fig:comparison_SecAgg_accessfl} abstracts the mechanisms of SecAgg compared to ACCESS-FL for one round of FL. This figure visualizes the process of generating masked models among four clients participating in a federated learning system. On the left side, representing SecAgg, we observe multiple clients (indicated by lightning bolt icons of different colors), each contributing to the FL process. These clients generate a masked model by combining three elements: their individual trained models' weights (shown by grid icons), shared masks created between themselves and every other client (illustrated by differently colored geometric shapes), and their own self-masks (shown as noise). As shown in the figure, as the number of clients increases, the number of shared masks each client needs to generate increases, leading to high communication and computation costs. Combining these elements results in a double-layered masking technique to conceal the trained models. Each client's masked model is then sent to a centralized server for aggregation. The server outputs an aggregated model (blue grid icon), which incorporates the knowledge learned from all participating clients without revealing any individual client's trained model and data. On the right side, ACCESS-FL is demonstrated. Each client still produces a trained model (grid icons) in this figure, but the masking process is simplified. Instead of creating a shared mask with every other client, each client only generates shared masks with two different clients (represented by the connection between the same colored geometric shapes and grid icons). As illustrated in the figure, the number of generated masks per client is independent of the network size; regardless of the number of clients, each client only needs to create shared masks with two other clients rather than all participants, which results in reduced message volume compared to SecAgg. These shared masks are added to the trained model to create a masked model, which is then sent to the server. The server aggregates these masked models into a new global model (blue grid icon). This diagram shows the contrast in complexity and message volume between the two protocols. SecAgg requires a more significant number of messages to be exchanged, as every client generates shared masks with every other client. ACCESS-FL, however, reduces the communication overhead by limiting the creation of shared masks with two clients, thereby reducing message volume and potentially increasing the overall efficiency of the FL process. The simplified masking process and reduced message volume in ACCESS-FL highlight its advantages over SecAgg in terms of communication and computation efficiency. \vspace{-2mm}
\subsection{Message Passing in ACCESS-FL}
This section analyzes the total number of messages exchanged between the server and $C$ number of clients over $n$ training FL rounds in ACCESS-FL. The process of messages passing in ACCESS-FL is categorized into three main phases as follows: 
\textbf{Phase $I$} (Initialization): The server \textbf{broadcasts} the initial model to all clients. Simultaneously, all clients receive common public parameters from a trusted third party. Then, each client generates a unique public-private key pair from public parameters and sends its public key to the server. Upon collecting all public keys, the server broadcasts a set of all public keys. \textbf{Phase $II$} (Shared mask generation at training round 1): Upon receiving public keys, each client calculates the index of two other clients in the participating list to create shared secrets using their public keys. Client $i$ calculates the index of its pairs as $fp_{i} = [(i+distance) \mod |C|]$ and $sp_{i} = [(i-distance+|C|) \mod |C|]$. Here, $distance$ is a random integer within the range of $[1, \left\lfloor \frac{|C|-1}{2} \right\rfloor]$ (where $|C| \geq 6$). We limit the upper domain to $\frac{|C|-1}{2}$ to make sure that the pairs are different; more than $\frac{|C|-1}{2}$ makes the chosen pair equal to the previously found pairs. After finding the pairs, the client generates shared secrets, which are used in a pseudo-random number generator that creates two shared masks (denoted as $m_{i, fp_{i}}$ and $m_{i, sp_{i}}$). \textbf{Phase $III$} (from training round 2): Each client $i$ performs model training and then computes the masked model as $w_{i}^{masked} = w_{i} + m_{i, fp_{i}} + m_{i, sp_{i}}$, where $w$ is the trained model. The computed $w_{i}^{masked}$ is sent to the server. Then, the server generates the new global model by aggregating all masked models. The masks cancel out each other due to the pairwise generation of shared secrets since $m_{i, fp_{i}} = - m_{fp_{i}, sp_{fp_{i}}}$ the sum of masked models equals the sum of unmasked trained models. Finally, the server broadcasts the new global model to all clients. Considering all communications after $n$ FL rounds, the total number of messages sent from all clients are $(n + 1) \times |C|$, and from the server are $n + 1$ messages. Considering all rounds, the communication order for each client and the server is $O(1)$. This analysis demonstrates the communication efficiency of ACCESS-FL, as the number of messages exchanged remains constant regardless of the network size.\vspace{-2mm}

\subsection{Explanation of Core Enhancements}
In this section, we explain the core enhancement of ACCESS-FL.

\noindent \textbf{Efficient key pair generation:}
In ACCESS-FL, one key pair is only required for creating shared secrets, whereas in SecAgg, two key pairs are necessary (one for shared secrets and one for encryption). Additionally, in ACCESS-FL, the key pair is generated once in the initial round. In contrast, in SecAgg, due to the reconstruction of self-masks for participants and the shared masks of dropout clients by the server, each client needs to generate a key pair in every round of FL. In SecAgg, if a client is delayed in sending its masked model, the server assumes it has dropped out. Consequently, every other client sends the portions of the delayed client's private key to the server, allowing the server to reconstruct the shared masks for this delayed client. If the delayed client's model is received in the following FL round, and the client continues to participate, the server receives portions of the client's random element and can calculate the self-mask. If the key pairs remain unchanged, the shared masks for this client also remain unchanged. After two rounds, the server can compute the client's trained model by subtracting the self-mask and shared masks from the masked model. Therefore, for security reasons, clients in SecAgg are required to generate new key pairs every round. However, in ACCESS-FL, clients do not share any information except for the masked model. This means the server cannot reconstruct the client's trained models due to the lack of private keys. Generating a key pair only once eliminates the need for clients to send their public keys to the server and for the server to broadcast these keys. This results in reduced communication and computation costs for both the server and clients. The efficient key pair generation in ACCESS-FL significantly reduces the computational burden on clients and the communication overhead between clients and the server.

\noindent \textbf{Simplified masking techniques:} ACCESS has refined the masking process to make it more communication and computation efficient. These enhancements include using a more compact representation of masks and employing less mathematical computation that requires fewer data to achieve the same level of privacy in honest-but-curious scenarios. In contrast to SecAgg, which uses both shared and self-masks, our enhanced protocol utilizes only shared masks. In Google's protocol, a double masking strategy is necessary because clients share secrets with the server. However, in our proposed method, applying self-masks is not required since we do not share any secrets with the server. The shared masks are generated between pairs of two participant clients. Each pair collaborates to create a masking vector with its peer from the pair client. By eliminating self-masks, our protocol significantly reduces the computational burden on each client. The focus on shared masks simplifies the entire mask generation process. Since each client is only responsible for generating and managing masks with two nodes, the overall complexity of the masking process is reduced. This masking approach reduces the computational cost and the amount of data that needs to be transmitted for masking purposes. The use of shared masks means fewer data packets are required to achieve the same level of privacy, leading to lower communication costs. The simplified masking techniques in ACCESS-FL lead to more efficient computation and communication compared to SecAgg's double-masking approach.\vspace{-1mm}

\section{Proof of Maintaining Aggregation Result Equal to Traditional FL} \label{sec: Proof of Maintaining Aggregation Result Equal to Traditional FL}

This section aims to demonstrate that in ACCESS-FL, the output of the aggregated model is maintained compared to traditional FL. The following settings for each client $i$ are considered: 1) $w_{i}$ represents the trained model of client $i$. 2) $m_{i, fp_{i}}$ represents the shared mask between client $i$ and its first paired client denoted by $fp_{i}$. 3) $m_{i, sp_{i}}$ represents the shared mask between client $i$ and its second paired client denoted by $sp_{i}$. 4) The equation $m_{i, j} = -m_{j, i}$ holds for any pair of clients $i$ and $j$ where $j < i$.

\noindent \textbf{Creating Masked Models:} The idea behind the proof is to use only two shared secrets to mask the individual models before aggregation. Each client $i$ creates a masked model $w_{i}^{masked}$ by adding its trained model $w_{i}$ with the shared masks $m_{i, fp_{i}}$ and $m_{i, sp_{i}}$. The equation $m_{i, j} = -m_{j, i}$ ensures that the shared masks cancel out when summed across all clients. This feature keeps the output of the aggregation function in ACCESS-FL equivalent to the output of the aggregation function in traditional FL.
Each client $i$ creates a masked model $w_{i}^{masked}$ as follows: 
\begin{align}
  w_{i}^{masked} = w_{i} + m_{i, fp_{i}} + m_{i, sp_{i}}.  
\end{align}

\noindent \textbf{Aggregating Masked Models:}
The aggregation of all masked models across $|C|$ clients, denoted as $W^{\text{masked}}$, is computed by summing up the masked models $w_{i}^{\text{masked}}$ for each client. This aggregate can be decomposed into three separate sums: the sum of the trained models $w_{i}$, the sum of the shared masks $m_{i, fp_{i}}$, and the sum of the shared masks $m_{i, sp_{i}}$, where $0 \leq i \leq |C|-1$. Thus,
{\small
\begin{align}
W_{aggregated}^{\text{masked}} &= \sum_{i=0}^{|C|-1} w_{i}^{\text{masked}}
&= \sum_{i=0}^{|C|-1} \left (w_{i} + m_{i, fp_{i}} + m_{i, sp_{i}} \right).
\end{align}}
We can decompose this into three separate sums:
\begin{align}
W_{aggregated}^{\text{masked}} &= \sum_{i=0}^{|C|-1} w_{i} + \sum_{i=0}^{|C|-1} m_{i, fp_{i}} + \sum_{i=0}^{|C|-1} m_{i, sp_{i}}.
\end{align}

\noindent \textbf{Cancellation of Shared Masks:}
Equation $m_{i, j} = -m_{j, i}$ implies that the shared mask generated by client $i$ with client $j$ is equal in magnitude but opposite in sign to the shared mask generated by client $j$ with client $i$. When these shared masks are summed across all clients, they cancel each other out. That is,
\begin{align}
\sum_{i=0}^{|C|-1} m_{i, fp_{i}} &= - \sum_{j=0}^{|C|-1} m_{j, sp_{j}}.
\end{align}
Thus, each shared mask $m_{i, fp_{i}}$ pairs with $m_{fp_{i}, sp_{fp_{i}}}$ such that:
\begin{align}
m_{i, fp_{i}} + m_{fp_{i}, sp_{fp_{i}}} &= 0.
\end{align}
Here, $sp_{fp_{i}}$ is defined as the second pair of the first pair of client $i$. Hence, summing across all clients:
\begin{align}
\sum_{i=0}^{|C|-1} m_{i, fp_{i}} + \sum_{i=0}^{|C|-1} m_{fp_{i}, sp_{fp_{i}}} &= 0.
\end{align}
and similarly, for the second paired client,
\begin{align}
\sum_{i=0}^{|C|-1} m_{i, sp_{i}} + \sum_{i=0}^{|C|-1} m_{sp_{i}, fp_{sp_{i}}} &= 0.
\end{align}
This means for every shared mask $m_{i, sp_{i}}$, there exists a corresponding $m_{sp_{i}, fp_{sp_{i}}}$ that cancels out. The notation $fp_{sp_{i}}$ represents the first paired client of the second pair of client $i$. Thus, the sum of shared masks $m_{i, fp_{i}}$ and $m_{i, sp_{i}}$ across all clients becomes zero. This cancellation property is a key feature of ACCESS-FL, ensuring that the shared masks do not affect the final aggregated model.

\noindent \textbf{Equivalence to Trained Models:} 
Therefore, the aggregation of all masked models $w_{i}^{\text{masked}}$ is equivalent to the sum of the individual trained models $w_{i}$ across all clients:
\begin{align}
W_{aggregated}^{\text{masked}} = \sum_{i=0}^{|C|-1} w_{i}.
\end{align}

The proof presented in this section establishes the correctness of ACCESS-FL, demonstrating that it achieves the same aggregation result as traditional FL while preserving the privacy of individual clients' models through the use of shared masks.

\section{Evaluation Result} \label{sec: Evaluation Result}
To verify the effectiveness of ACCESS-FL, we conduct experiments using the MNIST, FMNIST, and CIFAR10 datasets. MNIST is the main dataset in this paper, which contains 60,000 handwritten digit images used for training and 10,000 images for testing. Each image is a 28x28 gray-scale digit, and the goal is to classify the images into one of ten digit classes (0-9). In our experimental setup, we utilized 100 clients and assigned each client only 1 label to simulate a practical Non-Independent and Identically Distributed (Non-IID) scenario; We implement a 2-layer Neural Network (2NN) model consisting of an input layer for the flattened 28X28 pixel images, followed by two dense layers with 200 units and ReLU activation, and a final output layer of 10 units with softmax activation. Each experiments were conducted with 100 communication rounds with SGD \cite{bottou2010large} optimizer in a learning rate of 0.1. To assess the communication and computation costs of each protocol, we present results based on the accumulated message size, the number of exchanged messages, and the running time for both clients and the server in ACCESS-FL, SecAgg, and SecAgg+.

\subsection{Communication Cost of ACCESS-FL, SecAgg and SecAgg+}

\begin{figure}[t]
    \centering
    \begin{subfigure}[b]{0.23\textwidth}
        \includegraphics[width=1\columnwidth]{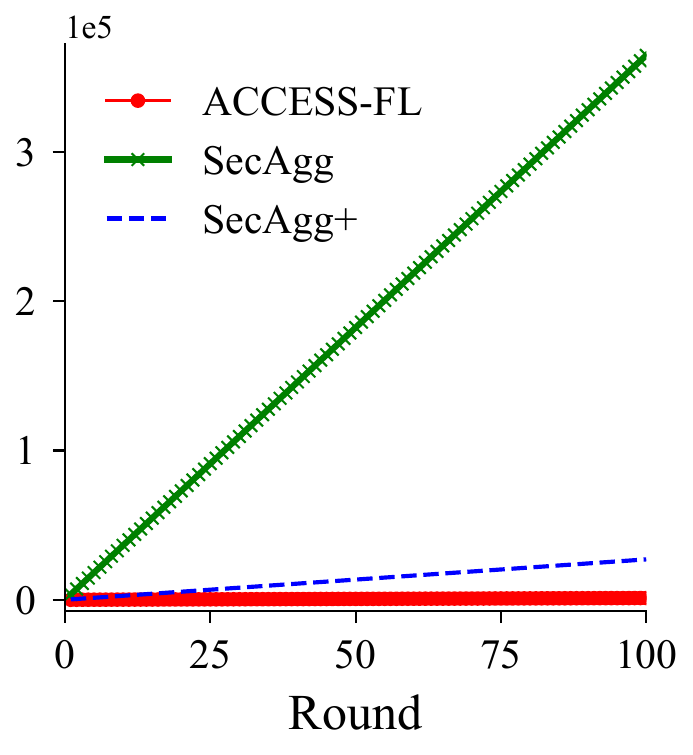}
        \vspace{-7 mm}
        \caption{Clients to server.}
        \label{fig:Client_Message_Size_kB}
    \end{subfigure}
    \begin{subfigure}[b]{0.23\textwidth}
        \includegraphics[width=1\columnwidth]{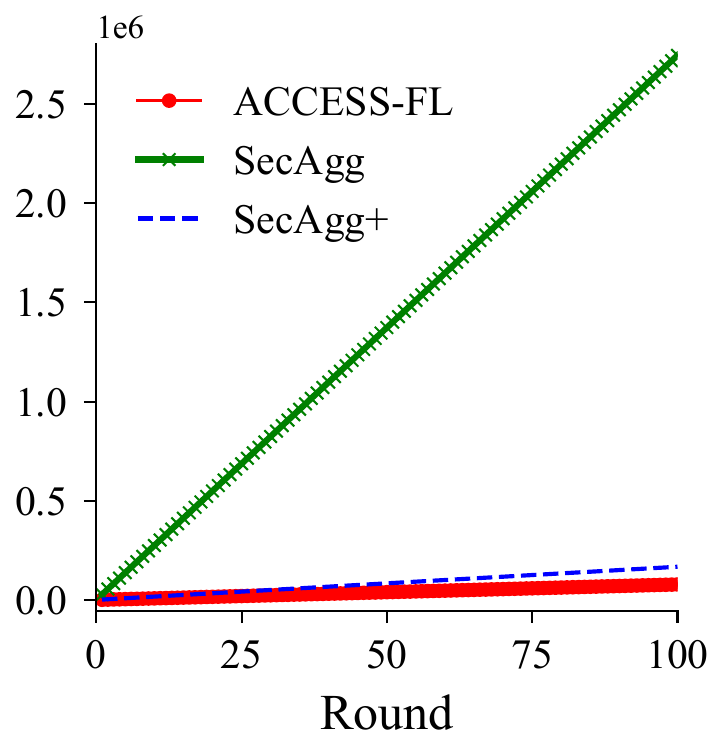}
        \vspace{-7 mm}
        \caption{Server to clients.}
        \label{fig:Server_Message_Size_kB}
    \end{subfigure}
    \vspace{-3 mm}
    \caption{Accumulative message size(kB) for MNIST experiments.}
    \vspace{-5 mm}
    \label{fig:accumulated_message_size}
\end{figure}

\begin{figure}[t]
    \centering
    \begin{subfigure}[b]{0.23\textwidth}
        \includegraphics[width=1\columnwidth]{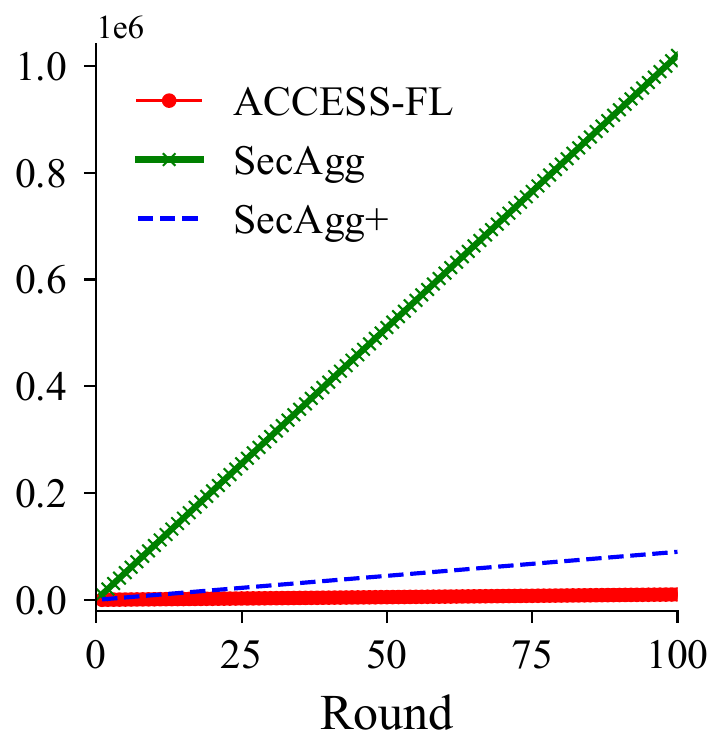}
        \vspace{-7 mm}
        \caption{Clients to server.}
        \label{fig:Client_Messages}
    \end{subfigure}
    \begin{subfigure}[b]{0.23\textwidth}
        \includegraphics[width=1\columnwidth]{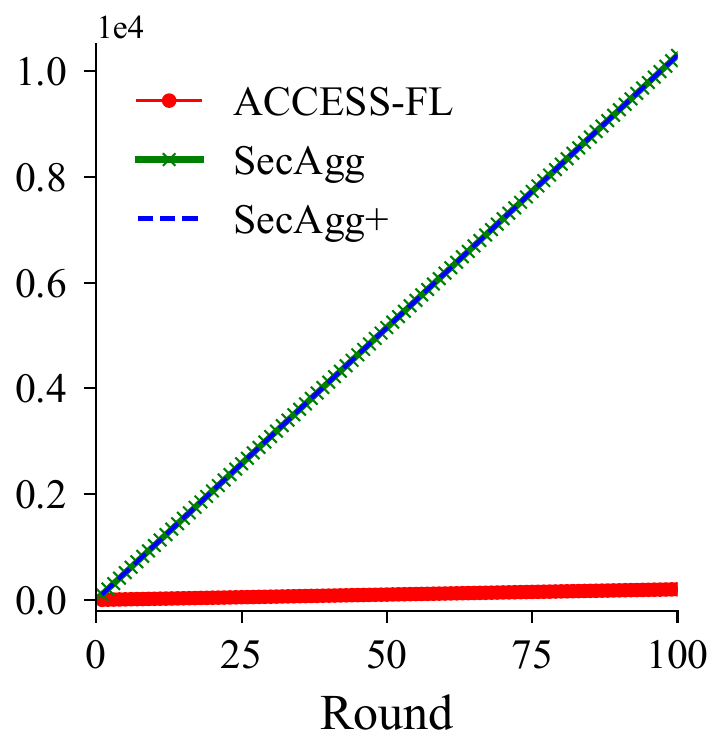}
        \vspace{-7 mm}
        \caption{Server to clients.}
        \label{fig:Server_Messages}
    \end{subfigure}
    \vspace{-3 mm}
    \caption{Accumulated number of messages for MNIST experiments.}\vspace{-4 mm}
    \vspace{-5 mm}
    \label{fig:accumulated_number_of_message}
\end{figure}

\begin{figure}[t]
    \centering
    \begin{subfigure}[b]{0.23\textwidth}
        \includegraphics[width=1\columnwidth]{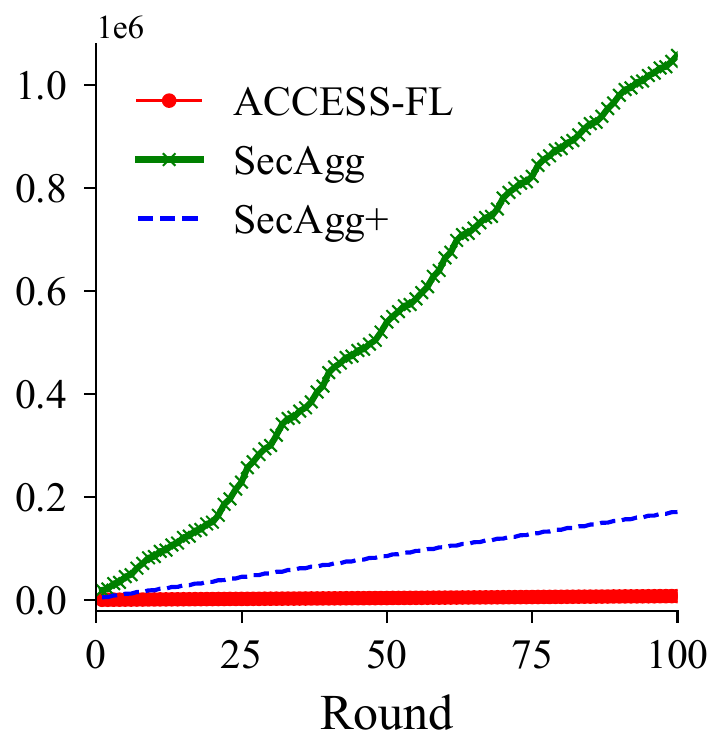}
        \vspace{-7 mm}
        \caption{Clients to server.}
        \label{fig:Client_Time_ms}
    \end{subfigure}
    \begin{subfigure}[b]{0.23\textwidth}
        \includegraphics[width=1\columnwidth]{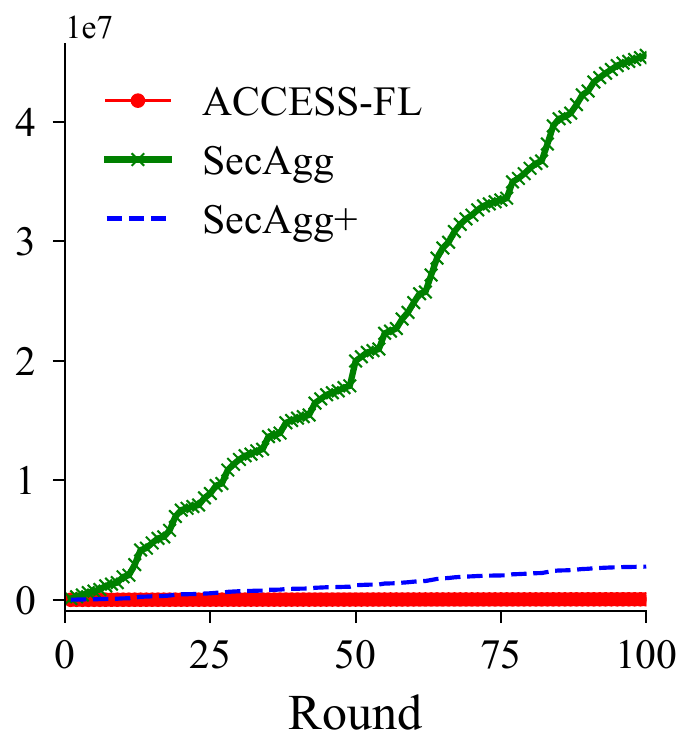}
        \vspace{-7 mm}
        \caption{Server to clients.}
        \label{fig:Server_Time_ms}
    \end{subfigure}
    \vspace{-3 mm}
    \caption{Accumulated running time on server and client(ms) for MNIST experiments.}
    \vspace{-5 mm}
    \label{fig:accumulated_running_time}    
\end{figure}

\begin{figure}[t]
  \centering
    \includegraphics[width=70 mm, height=50 mm]{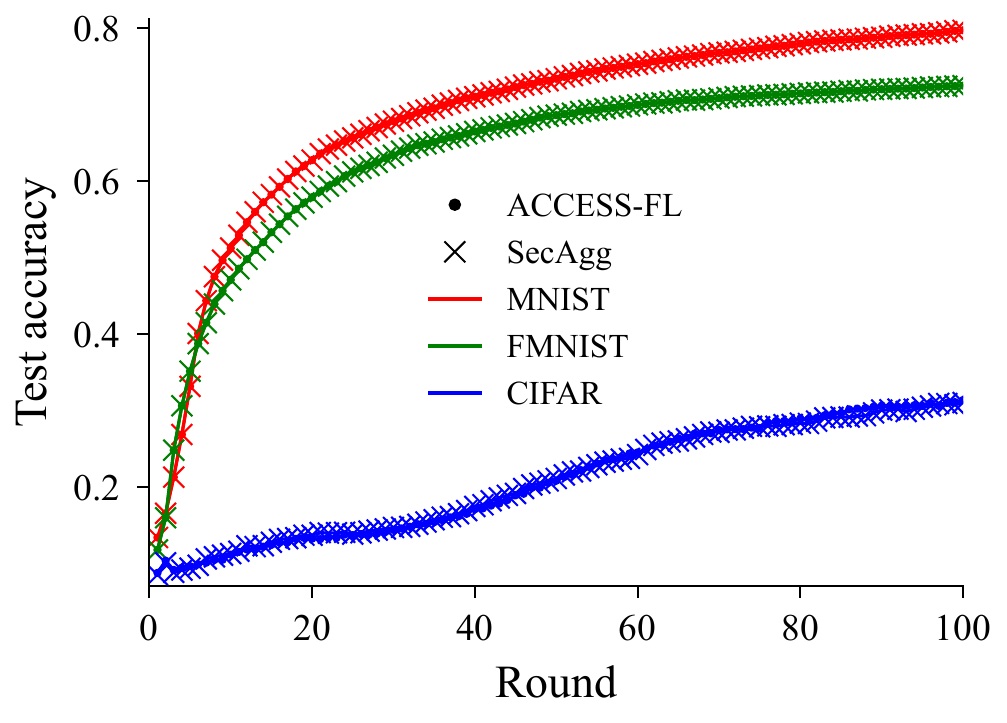}
    \vspace{-5 mm}
  \caption{Learning curve comparison between ACCESS-FL and FedAvg in MNIST, FMNIST and CIFAR.} 
  \label{fig:learning_curve}
\end{figure}

Figures \ref{fig:Client_Message_Size_kB}, \ref{fig:Server_Message_Size_kB} illustrate the accumulated message size sent from clients to server and sent from the server to clients, respectively, for protocols ACCESS-FL, SecAgg, and SecAgg+ over 100 rounds. We observe that the total size of the message for each client in ACCESS-FL remains at approximately 0.01 MB through the 100 rounds. And \textbf{the communication cost for each client does not increase with the number of participating clients}, as each client only generates shared masks with two pairs, regardless of the number of clients. In contrast, the total message size for each client in SecAgg and SecAgg+ increases with the number of clients because, in both algorithms, the number of pairs that each client generates depends on the network size. In SecAgg, each client pairs with every other client, and in SecAgg+, each client pairs with its $k$ neighbors in the predefined and randomly generated $k-regular$ graph by the server where $k = log(|C|)$ for the $|C|$ number of clients. In SecAgg, the total message size for each client is around 3.5MB by the 100th round. SecAgg+ reduces the message size for each client compared to SecAgg, but it still grows with the network size, reaching around 0.3MB after 100 training rounds. The server's accumulated message size in ACCESS-FL is almost 80MB by the 100th round. However, in SecAgg and SecAgg+, the volume of transmitted messages from the server through the network exceeds about 3000MB and 200MB, respectively, by the end of the 100 rounds. 

The accumulated number of messages exchanged between the server and clients is demonstrated in Figure \ref{fig:accumulated_number_of_message}. The number of messages sent by each client in ACCESS-FL stays constant at around 100 throughout the 100 rounds and demonstrates the scalability of ACCESS-FL, as the communication overhead on each client remains fixed regardless of the number of participating clients. On the other hand, the number of messages sent by each client in SecAgg for the network size of 100 clients is around 10,000 messages by the 100th round. The number of messages sent by each client in SecAgg+ decreases compared to SecAgg, approximately 900 messages by the 100th round. The number of messages sent by the server remains constant at 2 messages per round, except for the initial round, where an additional message is sent to broadcast the public keys. However, in SecAgg and SecAgg+, at each round, the server sends around 100 messages per round that include broadcasting two public keys per client, sending encrypted values received by every client, and broadcasting the participants in sending masked model and the new global model. Although the number of messages sent by the server is equal in both SecAgg and SecAgg+, the sizes of messages are considerably different. That is, in SecAgg, the server sends cipher texts to every client, which includes the encrypted value generated from every other client. However, in SecAgg+, the size of each encrypted-value message equals $log(|C|)$ as each client only pairs with its $log(|C|)$ neighbors. 

The constant communication cost for each client in ACCESS-FL is attributed to the protocol's design, which generates shared secrets between only two clients, regardless of the total number of participants. By limiting the number of pairwise shared secrets, ACCESS-FL significantly reduces the communication overhead for each client compared to SecAgg and SecAgg+. In ACCESS-FL, clients themselves \textbf{find their pairs and change them in every round without the server's knowledge.} The only message that each client sends at each round to the server is a masked model update (except for the initial round, where each client needs to generate one key pair and send its public key to the server). Hence, the number of transmitted messages from clients and the overall load on the network in ACCESS-FL is comparable to the number of messages that each client sends in traditional FL. Thus, ACCESS-FL improves the privacy of a hones-but-curious stable FL system with approximately the same load on clients that participated in a traditional FL. Additionally, the size of the masked model update remains constant in different network sizes, as the model architecture and the masking scheme do not change based on the number of participants. Furthermore, the number of messages sent from the server at each round is only twice compared to the traditional FL (except for broadcasting the received public keys at the first round). This message includes the new global model and the list of participants where the latter does not need any cryptographic operation. Thus, ACCESS-FL, in large-scale FL stable networks, makes the overhead on the server approximately equal to the server's overhead in traditional FL while prevents the server from applying a model inversion attack by concealing the trained model from a hones-but-curious server. In contrast, SecAgg and SecAgg+ require significantly higher communication costs due to the number of pairs that each client has and applying double masking along with the need for encryption to share secrets via server. In SecAgg, each client needs to generate shared secrets with every other client, which leads to an increase in the number of messages and message size for each client. SecAgg+, despite reducing the communication cost compared to SecAgg by having the clients pairing with $log(|C|)$ neighbors, still employs a double masking technique, and the clients are required to generate two key pairs and perform cryptographic operations to compute encrypted values for their neighbors. Despite this improvement, the server in SecAgg+ still knows who is paired with whom, and the communication cost for each client and the server grows with the number of clients. Furthermore, SecAgg+ requires message exchanges between the server and clients to handle the unmasking of the global model, even in stable FL environments with limited client dropouts and low delay variations. However, in ACCESS-FL, the server is not responsible for unmasking the aggregated masked models, and the protocol handles client dropout or incidents of delayed messages by making the clients find new pairs and resend their masked models. Which is efficient in large-scale stable FL systems such as healthcare systems where the privacy of data is crucial, the delay variation in the network is low, and the clients who participate in FL have reliable deployed devices, with a rare client dropout rate. Additionally, the server in such networks is honest-but-curious; However, it is required to prevent the server from accessing the critical clinical data of patients. Applying SecAgg and SecAgg+ for such networks makes unnecessary overload on the network, clients, and server only for privacy-preserving.

Figure \ref{fig:Server_Time_ms} illustrates the accumulated running time for the server in ACCESS-FL, SecAgg, and SecAgg+ over the 100 training rounds. In ACCESS-FL, the server's running time at each FL round is constant (except for the initial round, where the server broadcasts the public keys to the network) and approximately reaches accumulatively 30 seconds by the end of the 100 rounds. This computation cost consists of the server aggregating the masked model updates without being responsible for handling client dropout or delayed messages, in contrast to SecAgg and SecAgg+, the server is required to unmask the aggregated model and manage the client dropout. Upon receiving masked models from all participating clients, the server aggregates these updates, and generates the new global model in each round. As the number of clients remains constant (100 in our experiments), the server's running time per round also remains relatively constant. On the other hand, the server's accumulated running time in SecAgg increases quadratically, exceeding 445,000 seconds by the 100th round and approximately 2 minutes per round. SecAgg+ reduces the server's running time compared to SecAgg, summing up to around 3000 seconds by the 100th round. The high computation costs on the server side in SecAgg and SecAgg+ are the result of several factors, such as the need to perform cryptographic operations to reconstruct the shared masks for dropped-out clients and self-masks for participants (by using Shamir's secret sharing to generate the random elements of participants and private key of dropped-out clients, then running PRG function on the calculated elements to reconstruct the masks) that leads to an increased complexity of the aggregation process. Although SecAgg+  introduces an additional computation cost for the server to generate the random graph, the overall server cost is lower than SecAgg because the number of pairs per client is reduced from $|C|-1$ in SecAgg (where each client is peered with every other client) to $log(|C|)$ in SecAgg+.

Figure \ref{fig:Client_Time_ms} shows the accumulated running time for clients in ACCESS-FL, SecAgg, and SecAgg+ over the 100 training rounds. In ACCESS-FL, the running time for each client remains low and does not increase with the number of participating clients, staying at approximately 0.4 seconds throughout the 100 rounds. The constant computation cost for each client demonstrates that the protocol remains suitable for large-scale stable networks. In ACCESS-FL, each client runs a deterministic function to find its two pairs, performs local model training, and applies the masking process to the trained model (by using the shared secrets generated with only two pairs and running PRG function on the shared secrets to generate the shared masks), then it sends the masked model update to the server. In case of client dropout or delayed message, the server sends the participants within the same training round, then clients find new pairs and mask their trained models with the new shared masks. Thus, ACCESS-FL only applies dropout mitigation techniques or handles delayed updates in ACCESS-FL only when necessary.
In contrast, the running time for each client in SecAgg is about 10 seconds by the 100th round. SecAgg+ reduces the running time for each client compared to SecAgg, but it still increases with the number of clients (approximately 2 seconds by the 100th round. The computation cost for each client in SecAgg and SecAgg+ is a consequence of their more complex masking processes, which involve creating two key pairs, generating shared secrets (with every other client in SecAgg and the $log(|C|)$ neighbors in SecAgg+), performing double masking to its trained model, running PRG function on the random element to generate the self-mask, running $|C|-1$ times of PRG function in SecAgg and $log|C|$ in SecAgg+ on shared secrets to creating shared masks, engaging in cryptographic operations that include splitting their private key and random element into $|C|$ parts in SecAgg and $|log(|C|)$ parts in SecAgg+, then it encrypts these values with its pair public key and sends these cipher texts to the server. After the server receives the masked models and sends the participants list, each client needs to decrypt the cipher texts of its pairs. Then, it sends the portion of the random element of its pairs if they are claimed as a participant by the server or the portion of the private key of its peers if they are recognized as the dropped-out clients by the server. Lastly, Figure \ref{fig:learning_curve} shows the comparison of learning curves between ACCESS-FL and FedAvg on the MNIST, FMNIST, and CIFAR datasets. When ACCESS-FL is applied, the results between ACCESS-FL and FedAvg on MNIST and FMNIST are exactly the same, and the accuracy difference on CIFAR is less than 1\% in each training round. 
\vspace{-3mm}
\subsection{Client Dropout for ACCESS-FL, SecAgg, and SecAgg+}

\begin{table*}[t]
\centering
\begin{tabular}{|c|c|c|c|c|c|c|c|}
\hline
\textbf{Round} & \textbf{ACCESS-FL (ND)} & \textbf{ACCESS-FL (D)} & \textbf{SecAgg (ND)} & \textbf{SecAgg (D)} & \textbf{SecAgg+ (ND)} & \textbf{SecAgg+ (D)} & \textbf{FedAvg (ND)} \\ \hline
10  & 1100  & 1099  & 102000   & 111702  & 9000   & 9594   & 1000\\ \hline
30  & 3100  & 3067  & 306000   & 4312520 & 27000  & 379764 & 3000\\ \hline
50  & 5100  & 4995  & 510000   & 12592570 & 45000  & 1109910 & 5000\\ \hline
70  & 7100  & 6883  & 714000   & 24951868 & 63000  & 2200032 & 7000\\ \hline
100 & 10100 & 9640  & 1020000  & 51139440 & 90000  & 4510170 & 10000\\ \hline
\end{tabular}
\caption{Total number of messages sent from clients for scenarios with \textbf{node dropout (D) and without node dropout (ND)}.}\vspace{-7mm}
\label{tab:client_messages}
\end{table*}

\begin{table*}[t]
\centering
\begin{tabular}{|c|c|c|c|c|c|c|c|}
\hline
\textbf{Round} & \textbf{ACCESS-FL (ND)} & \textbf{ACCESS-FL (D)} & \textbf{SecAgg (ND)} & \textbf{SecAgg (D)} & \textbf{SecAgg+ (ND)} & \textbf{SecAgg+ (D)} & \textbf{FedAvg (ND)} \\ \hline
10  & 12   & 13   & 1030   & 1228   & 1030   & 2218   & 11\\ \hline
30  & 32   & 35   & 3090   & 43848  & 3090   & 46788  & 31\\ \hline
50  & 52  & 57  & 5150   & 127660 & 5150   & 132510 & 51\\ \hline
70  & 72  & 79  & 7210   & 252664 & 7210   & 259384 & 71\\ \hline
100 & 102  & 112  & 10300  & 517405 & 10300  & 526855 & 101\\ \hline
\end{tabular}
\caption{Total number of messages sent from the server for scenarios with \textbf{node dropout (D) and without node dropout (ND).}}\vspace{-7mm}
\label{tab:server_messages}
\end{table*}

\begin{table*}[t]
\centering
\begin{tabular}{|c|c|c|c|c|c|c|c|}
\hline
\textbf{Round} & \textbf{ACCESS-FL (ND)} & \textbf{ACCESS-FL (D)} & \textbf{SecAgg (ND)} & \textbf{SecAgg (D)} & \textbf{SecAgg+ (ND)} & \textbf{SecAgg+ (D)} & \textbf{FedAvg (ND)} \\ \hline
10  & 8.58   & 8.58   & 274.620  & 274.639  & 16.722  & 16.838  & 8.56\\ \hline
30  & 24.143  & 24.144  & 823.859  & 11534.090 & 50.165  & 702.659 & 24.123\\ \hline
50  & 39.707  & 39.708  & 1373.100 & 33778.320 & 83.609  & 2057.346 & 39.686\\ \hline
70  & 55.27  & 55.272  & 1922.338 & 67007.340 & 117.052 & 4080.900 & 55.25\\ \hline
100 & 78.615  & 78.619  & 2746.197 & 137447.400 & 167.218 & 8370.354 & 78.594\\ \hline
\end{tabular}
\caption{Total size of messages sent from the server (MB) for scenarios with \textbf{node dropout (D) and without node dropout (ND).}}\vspace{-7mm}
\label{tab:server_message_size}
\end{table*} \vspace{-2mm}

In this section, we compare the communication costs of ACCESS-FL, SecAgg, and SecAgg+ under scenarios with and without client dropout. We evaluate the number of messages sent from clients and servers, as well as the size of these messages over 100 rounds on MNIST dataset. Table \ref{tab:client_messages} presents the comparison of the number of messages sent from clients. In the ACCESS-FL, the number of messages sent from clients in the dropout scenario slightly decreases compared to the stable scenario, with the reduction becoming more pronounced as the rounds progress. For instance, at the 100th round, the number of messages drops from 10,100 to 9,640 due to client dropouts. This is because, in each dropout scenario, the remaining clients compensate by sending additional portions of the dropout client’s shares to the server. In contrast, SecAgg shows a significant increase in the number of messages in dropout scenarios due to the overhead of handling client dropouts and reconstructing shared secrets. SecAgg+ also shows an increase, though it is less severe compared to SecAgg, due to its more efficient handling of client pairs in a k-regular graph structure. Table \ref{tab:server_messages} illustrates the number of messages sent from the server. ACCESS-FL demonstrates a constant and minimal increase in the number of server messages, as it maintains a steady communication pattern irrespective of client dropouts. In SecAgg and SecAgg+, the server’s messaging overhead significantly increases in the presence of dropouts, reflecting the additional communication required to manage shared secret reconstructions and handle the redistribution of keys and masked values. Table \ref{tab:server_message_size} provides the size of messages sent from the server. ACCESS-FL maintains a smaller message size, around 78.61 MB at the 100th round, even in the dropout scenario. This reveals the protocol’s efficiency in managing communication overhead to handle client dropouts. In contrast, SecAgg’s message size reaches 137.45 GB due to the intensive cryptographic operations required to manage dropouts and double masking. Although SecAgg+ reduces this overhead, it grows substantially in message sizes and reaches approximately 8.37 GB by the 100th round. The comparative analysis of ACCESS-FL with SecAgg and SecAgg+ demonstrates that ACCESS-FL significantly reduces both the number and size of messages exchanged between clients and the server. This reduction is achieved by eliminating unnecessary cryptographic operations and having shared secrets only between two other peers per client. ACCESS-FL is particularly effective in stable federated learning environments with limited client dropout rates and low network frequencies. \vspace{-2mm}

\section{Related Work} \label{sec: Related Work}
Different papers have worked on secure aggregation for FL aimed at reducing communication and computation costs. Authors in \cite{mandal2018nike} proposed a non-interactive key establishment protocol that eliminates Shamir's secret sharing to reduce both communication and computation overheads. Furthermore, FastSecAgg \cite{kadhe2020fastsecagg} employs a multi-secret sharing scheme based on Fast Fourier Transform (FFT \cite{heckbert1995fourier}) to achieve significantly lower computation costs while maintaining similar communication costs as SecAgg. Addressing communication overhead, the SAFER method \cite{beguier2020safer} compresses the neural network updates by using the TopBinary Coding and one-bit quantization to reduce the data sent during training. In a different approach, SAFELearn \cite{fereidooni2021safelearn} presented a flexible secure aggregation protocol that is adaptable to various security and efficiency demands. SAFELearn can be implemented with the Full HE (FHE) \cite{albrecht2015ciphers}, Multi-Party Computation (MPC), or the Secure Two-Party Computation (STPC) to protect privacy. The main features of the SAFELearn system include the need for only two rounds of communication in each training iteration, the ability to handle client dropouts, and avoiding reliance on any trusted third party. The work of Wu et al. \cite{wu2024security} critically examined the Verifiable and Oblivious Secure Aggregation (VOSA) protocol and demonstrated vulnerabilities that could allow forgery attacks by a malicious aggregation server. Complementing this, Mansouri et al. in\cite{mansouri2023sok} offered a systematic evaluation of secure aggregation protocols for FL and categorized them by cryptographic schemes. They identified challenges such as client failures, inference attacks, and malicious activities.
Further addressing security enhancements, Rathee et al. \cite{rathee2023elsa} introduced ELSA, a protocol designed to counter active adversaries with improved efficiency. ELSA reduces communication and computation costs while maintaining privacy integrity in the presence of malicious actors. Liu et al. \cite{liu2022efficient} presented a scalable privacy-preserving aggregation scheme that addresses honest-but-curious and active adversaries and introduced dropout resilience.

Georgieva et al. in \cite{georgieva2023falkor} proposed the Falkor protocol for secure and efficient aggregation by using GPU acceleration and stream cipher-based masking. This approach scales efficiently across multiple servers and enhances privacy without compromising computational efficiency. Gupta et al. \cite{gupta2023resource} targeted the specific needs of urban sensing systems with their Resource Adaptive Turbo-Aggregate protocol to showcase adaptability to varying network resources, which is practical in real-world application of FL in resource-constrained environments. Pejo et al. \cite{pejo2023quality} investigated the quality inference challenge within FL and proposed scoring rules to evaluate participants' data contributions. Their methodology enhances model training efficiency and enables the detection of misbehaving participants, as a critical aspect of collaborative learning environments. Authors in \cite{jegadeesan2023blockchain} explored the application of blockchain in smart farming and proposed a blockchain-based aggregation that improves data management and productivity while incorporating IoT technologies for a smart agriculture system.

The threat of model poisoning is addressed by Wang et al. \cite{wang2023robust} through the Client Selection Secure Collaborative Learning (CSSCL) algorithm, which utilizes similarity metrics to ensure the integrity of model aggregations. This method represents a critical defense mechanism against the potentially harmful impacts of malicious clients on collaborative learning systems. In the area of edge computing, Wang et al. \cite{wang2020secure} proposed a Blockchain-based Secure Data Aggregation strategy (BSDA), which employs a novel security label system to ensure task integrity and confidentiality to enhance data aggregation methods in IoT networks. Also, Bouamama et al. \cite{bouamama2023edgesa} designed EdgeSA for privacy-preserving FL in edge computing environments. Their use of pairing-based cryptography and decentralized key generation addresses privacy concerns and resource constraints of edge devices to apply the secure aggregation approach in edge environments.

Authors in \cite{elkordy2023much} provided formal privacy for FL with existing secure aggregation protocols. They theoretically quantify the privacy leakage in FL when using secure aggregation with the FedSGD \cite{mcmahan2017communication} protocol. They derive upper bounds on how much information about each user's dataset can leak through the aggregated model update, using Mutual Information (MI) \cite{kraskov2004estimating} as the quantification metric. Their theoretical bounds show that when using the FedSGD aggregation algorithm, the amount of privacy leakage reduces linearly with the number of users participating in FL with secure aggregation. They use an MI Neural Estimator to empirically evaluate the privacy leakage under different FL setups on the MNIST and CIFAR10 datasets. Their experiments show a reduction in privacy leakage as the number of users and local batch size grow and an increase in privacy leakage as the number of training rounds increases. Moreover, they observe similar empirical dependencies of privacy leakage on FL parameters for the FedAvg and FedProx \cite{li2020federated} protocols.

\section{Discussion and Future Work} \label{sec: Discussion and Future Work}
ACCESS-FL is optimized for large-scale, stable FL environments where node dropout is limited and network delays are low. Practical implementations of such environments include fraud detection for financial applications \cite{yang2019ffd}, privacy-preserving systems against money laundry by IBM\cite{ibm2023privacy}, and AI applications in healthcare systems \cite{rahman2023federated}. These applications could benefit from reduced communication and computation overhead, which makes ACCESS-FL a practical choice for privacy-sensitive domains.
Future work could involve extending ACCESS-FL to handle active adversaries. Additionally, the integration of differential privacy techniques with ACCESS-FL could further enhance the privacy guarantees of the protocol. However, a limitation of ACCESS-FL is its performance when node dropout or delayed messages occur frequently, as this can lead to loop vulnerability where clients are stuck within a training round while finding new pairs.\vspace{-3mm}

\section{Conclusion} \label{sec: Conclusion}
In this paper, we proposed ACCESS-FL, an efficient, secure aggregation protocol designed for honest-but-curious scenarios in a stable FL environment with limited client dropout and low network delay variations. ACCESS-FL addresses the high communication and computation costs associated with Google's SecAgg protocol and SecAgg+ while maintaining the same level of security against model inversion attacks. ACCESS-FL generates shared secrets between only two clients, regardless of the number of clients, which reduces the computational complexity to a constant level and makes the communication cost for each client $O(1)$. Our protocol eliminates the need for double-masking, cryptographic computations, and self-masks by having only shared masks which cancel out each other during the aggregation process without server intervention. This approach significantly reduces the computational and communication burden on both clients and servers. ACCESS-FL handles client dropouts or delayed updates by having participating clients generate new shared masks with new peers and resend their masked models, which ensures the server is not required to manage the removal of masks from dropped-out clients. We conducted experiments on the MNIST dataset to evaluate the performance of ACCESS-FL compared to SecAgg and SecAgg+. The evaluation results demonstrated that ACCESS-FL significantly reduces communication and computational costs. The accumulated message size and number of messages exchanged between the server and clients remained constant in ACCESS-FL, whereas they increased with the number of clients in SecAgg and SecAgg+. Furthermore, the running time for the server and clients in ACCESS-FL was substantially lower than in SecAgg and SecAgg+.\vspace{-2.5mm}

\bibliographystyle{splncs04}
\bibliography{mybibliography}
\appendix
\end{document}